\documentclass[aps,prb,reprint,superscriptaddress,amsmath,amssymb]{revtex4-2}
\usepackage[utf8]{inputenc} % Add this line

\DeclareUnicodeCharacter{0307}{} % Add this line if necessary
\usepackage{xcolor}
\usepackage{graphicx}% Include figure files
\DeclareGraphicsExtensions{.png,.pdf,.tif,.jpeg}
\DeclareUnicodeCharacter{03BA}{\ensuremath{\kappa}}
\usepackage{parskip}  
\usepackage{siunitx}
\usepackage{amsmath}
\usepackage{hyperref}
\hypersetup{colorlinks,allcolors=black}
\begin{document}

% Use the \preprint command to place your local institutional report
% number in the upper right-hand corner of the title page in preprint mode.
% Multiple \preprint commands are allowed.
% Use the 'preprintnumbers' class option to override journal defaults
% to display numbers if necessary
%\preprint{}

%Title of paper
\title{Coexistence of Nodal and Nodeless Pairing Symmetry in Superconducting 6R-SnNbSe$_2$}
% repeat the \author .. \affiliation  etc. as needed
% \email, \thanks, \homepage, \altaffiliation all apply to the current
% author. Explanatory text should go in the []'s, actual e-mail
% address or url should go in the {}'s for \email and \homepage.
% Please use the appropriate macro foreach each type of information

% \affiliation command applies to all authors since the last
% \affiliation command. The \affiliation command should follow the
% other information
% \affiliation can be followed by \email, \homepage, \thanks as well.
\author{K. Yadav}
\author{M. Lamba}
\author{S. Patnaik}
\email{spatnaik@jnu.ac.in}
\affiliation{School of Physical Sciences, Jawaharlal Nehru University, New Delhi-110067, India}

%Collaboration name if desired (requires use of superscriptaddress
%option in \documentclass). \noaffiliation is required (may also be
%used with the \author command).
%\collaboration can be followed by \email, \homepage, \thanks as well.
%\collaboration{}
%\noaffiliation

\date{\today}
\begin{abstract}

Majorana fermions, a fundamental idea to fault-tolerant quantum computing, can emerge in systems where superconductivity coexists with nontrivial band topology. One promising route to realizing such topological superconductors (TSCs) involves inducing superconductivity in topological materials, particularly in systems lacking inversion symmetry. In this study, we report the synthesis and detailed characterization of Sn-intercalated NbSe\textsubscript{2}, forming a new polytype, 6R-SnNbSe\textsubscript{2}. This compound crystallizes in the noncentrosymmetric space group $R\overline{3}m$ and exhibits bulk superconductivity below $T_c \approx 4$~K. Structural, electronic, and magnetic measurements confirm the emergence of a superconducting phase derived from Sn intercalation into the non-superconducting 3R-NbSe\textsubscript{2}. Temperature-dependent magnetic penetration depth and superfluid density measurements down to 1.5~K are performed using the tunnel diode oscillator technique. The findings suggest the mixing of nodal and nodeless superconductivity in 6R-SnNbSe\textsubscript{2}. Given the noncentrosymmetric nature of the crystal structure and the theoretical prediction of topological nodal-line features in SnNbSe\textsubscript{2}, it is an interesting candidate to investigate unconventional pairing mechanisms. Our findings highlight the potential of this material to host nontrivial superconducting states among the transition-metal dichalcogenides.

\begin{description}
\item[PACS numbers] 74.25.Fy, 74.25.Ha, 74.25.Qt, 74.50.+r
\item[Keywords]
topological superconductivity, penetration depth, superfluid density
\end{description}
\end{abstract}

%\keywords{Suggested keywords}%Use showkeys class option if keyword
                              %display desired
\maketitle

%\tableofcontents

\section{Introduction}
The material realization of topological superconductivity accrues great current relevance because of its potential usage as qubits in fault-tolerant quantum computation~\cite{lutchyn2018majorana, nayak2008non, sato2017topological}. Towards this end, two approaches have been prescribed: one is the proximitized $s$-wave superconductivity on spin helical states~\cite{fu2008superconducting, nadj2014observation, jack2019observation, sun2016majorana}, and the other is to realize an intrinsic topological superconductor with odd-parity spin-triplet pairing~\cite{mackenzie2003superconductivity, joynt2002superconducting, matano2016spin, zhang2018observation, liang2021three}. Analogous to a topological insulator, in a three-dimensional topological superconductor (TSC), the superconducting bulk gap of the interior coexists with gapless surface states \cite{liang2021three, zhang2022topological}. The quasiparticle excitations of these surface states are theorized as Majorana fermions, and the experimental manifestation of such an exotic field-theoretical perspective is considered an essential component for the characterization of TSC. In this letter, we report such a possibility in superconducting 6$R$-SnNbSe$_2$ by determining its order parameter symmetry from temperature dependent penetration depth measurements. \par

One of the most promising approaches to achieving TSCs is to induce superconductivity in topological materials through doping or applying external pressure \cite{Wray2010Observation, Qi2016Topological, He2016Pressure-induced}. Bi$_2$Se$_3$ and its doped variants, such as Cu$_x$Bi$_2$Se$_3$, Sr$_x$Bi$_2$Se$_3$, and Nb$_x$Bi$_2$Se$_3$, have been extensively studied. They exhibit various phenomena indicative of unconventional pairing, including zero-bias conductance peaks, breaking of time-reversal symmetry, and nematic superconductivity \cite{yonezawa2018nematic, neha2019time, shen2017nematic}. Despite these findings, the pairing symmetry and exact character of the superconductivity for application as qubits has remained under debate. \par

In topological superconductivity, the physical properties depend on material symmetries. Time-reversal, spin-rotation, and inversion symmetries collectively determine topological characteristics \cite{Schnyder2008Classification}. A topological superconducting state emerges when odd-parity pairing coincides with Fermi surfaces that enclose an odd number of time-reversal-invariant momenta \cite{sato2010topological}. Noncentrosymmetric superconductors (NCS), where the crystal structure lacks inversion symmetry, present a platform for such possibilities \cite{naskar2021superconductors}. In NCS systems, antisymmetric spin-orbit coupling of Rashba type leads to mixing of spin-singlet and spin-triplet pairing components in the superconducting order parameter \cite{bauer2012non, smidman2017superconductivity}. This mixing of nodal and nodeless characters typically occurs in materials that lack inversion symmetry and exhibit strong spin-orbit coupling. The resulting superconducting states can host Majorana-bound states in vortex cores and gapless surface modes that is a hallmark of nontrivial topological phases \cite{santos2010superconductivity, sato2009topological, chang2014majorana}. \par

In this regard, SnNbSe$_2$ is considered as a promising material that theoretically combines nodal-line semimetal behavior with superconductivity. First principle studies predict that SnNbSe$_2$ is a superconductor with nontrivial $\mathbb{Z}_2$ topology and a transition temperature approximately 7~K \cite{chen2016ab}. It belongs to the same family as PbTaSe$_2$, a well-known noncentrosymmetric superconductor with topological nodal lines confirmed by ARPES and first-principles calculations \cite{wilson2017mu, chang2016topological}. In these materials, the absence of inversion symmetry and strong SOC create conditions favorable for spin-triplet pairing and unconventional superconducting gap structures. \par

In this study, we report the synthesis and characterization of a new polytype, 6R-SnNbSe$_2$, derived from Sn intercalation into 3R-NbSe$_2$. While previous theoretical predictions suggested a superconducting phase with space group $P6m2$ (No.~187), our experimental findings based on structural, electronic, and magnetic measurements reveal superconductivity in 6R-SnNbSe$_2$ belonging to the $R\overline{3}m$ space group (No.~166). We present measurements of $\Delta\lambda(T)$ down to 1.5~K in polycrystalline SnNbSe$_2$, where lower temperatures and higher resolution permit more precise determination of superconducting pairing symmetry. Our findings indicate the potential coexistence of nodal and nodeless pairing, marking that 6R-SnNbSe$_2$ may support topologically nontrivial superconducting states driven by the interplay of broken inversion symmetry and spin-orbit coupling.

\begin{figure*}
\includegraphics[width=1\textwidth,height=6cm]{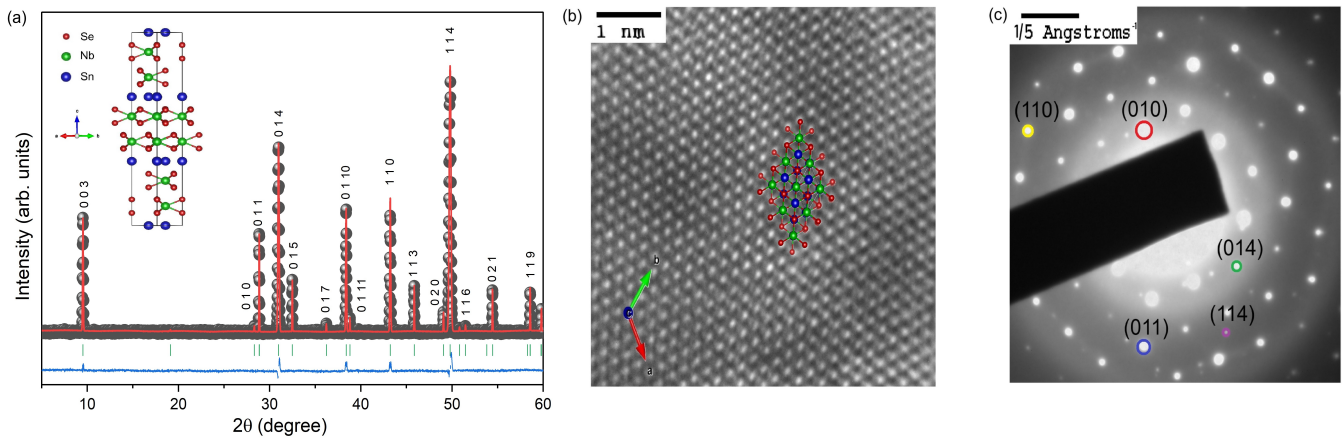}
  \caption{(a) Rietveld refinement of PXRD pattern of polycrystalline 6R-SnNbSe$_2$ and inset shows crystal structure of the unit cell. (b)  HRTEM image of 6R-SnNbSe$_2$ with atomic resolution. (c) SAED pattern with calculated Miller indices.}
\end{figure*}

\section{EXPERIMENTAL TECHNIQUES}
Polycrystalline SnNbSe\textsubscript{2} was synthesized using a three-step solid-state reaction method. The high-purity niobium (Alpha Aesar, 99.9\%) and selenium (Sigma Aldrich, 99.99\%) were accurately weighed in the stoichiometric ratio. It was subsequently ground for 30 minutes to homogenize the mixture and then pressed into pellet form using a hydraulic press. The pellet was vacuum-sealed ($10^{-2}$ to $10^{-3}$~mbar) in a quartz tube to avoid oxidation. The sample was then annealed at $850\,^\circ$C for 48 hours in a muffle furnace to yield NbSe\textsubscript{2}. For the second step, NbSe\textsubscript{2} was mixed with a stoichiometric amount of tin (Sn), re-ground for 30 minutes, and pressed into pellets. The pellets were sealed again in a quartz tube and then annealed at $850\,^\circ$C for 48 hours to yield polycrystalline SnNbSe\textsubscript{2}. For the last step, as grown SnNbSe\textsubscript{2} was powdered, pelletized, and then sintered at $800\,^\circ$C for 24 hours, followed by slow cooling.\par

Crystallographic orientation and phase purity were affirmed by room temperature X-ray diffraction (XRD) with a Rigaku Miniflex-600 diffractometer having Cu-$K\alpha$ radiation ($\lambda = 1.54056\,~$\AA). Surface morphology and elemental distribution were examined by scanning electron microscopy (SEM). Energy-dispersive X-ray analysis (EDX) was performed on a Zeiss EVO40 SEM analyzer and a Bruker AXS microanalyzer. Additional structural analysis was performed using high-resolution transmission electron microscopy (HRTEM) on a JEOL TEM, allowing precise lattice fringe and defect structure imaging. Electrical and magneto-transport properties were carried out under various magnetic fields and temperatures using a cryogen-free magnet (\textit{Cryogenic}, 8~T). Further, the temperature-dependent magnetization measurements were conducted using a vibrating sample magnetometer (VSM) probe in a physical property measurement system (\textit{Cryogenic}, 14~T).

The penetration depth was measured using the tunnel diode oscillator (TDO) technique, a highly sensitive method that utilizes stable high-frequency LC oscillations, typically in the MHz range. This technique detects subtle changes in the superconducting state by measuring shifts in the resonance frequency of the oscillator, which is directly linked to variations in the magnetic penetration depth. In this setup, the superconducting sample is placed within a small coil that serves as the inductor in a resonant LC circuit driven by a radio-frequency (RF) oscillator. The RF field generated by the coil is much weaker (in the $\mu$T range) than the sample's lower critical field \( H_{c1} \). Changes in the oscillator's resonance frequency with temperature are indicative of increased flux within the sample, which can be directly correlated to the penetration depth of the material. \par

\section{RESULTS AND DISCUSSION}

  FIG.~1.~(a) shows the Rietveld refinement of PXRD data for polycrystalline SnNbSe\textsubscript{2}, analyzed using Fullprof software. All reflections match with $R\overline{3}m$ space group. While NbSe\textsubscript{2} forms a 3R polytype due to its trilayer structure exhibiting the $R3m$ space group, Sn intercalation induces a structural transformation to the 6R polytype. The inset of FIG.~1.~(a) illustrates the detailed crystallographic structure of SnNbSe\textsubscript{2} (space group 166, $R\overline{3}m$) with a rhombohedral stacking. Each stacking unit consists of a Sn layer and a bilayer of NbSe\textsubscript{2}, which is rotated 60° relative to each other. This stacking leads to six-layer periodicity defining the 6R structure. Similar structural transitions are observed in TMDs under external or chemical pressure \cite{xiao2021structural, xiao2022pressure, liu2021cobalt, kasinathan2009afe2as2, kimber2009similarities, yao2016enhanced}. For instance, Rb doping in NbSe\textsubscript{2} induces a 2H to 6R phase transition with electronic modifications \cite{fan2019effects}. Likewise, 2H-Cu\textsubscript{x}Ta\textsubscript{1+y}S\textsubscript{2} transforms into 6R polytype under slow cooling during synthesis \cite{antal2017influence}. The calculated lattice parameters for SnNbSe\textsubscript{2} are $a = b = 3.661$~\AA{} and $c = 27.831$~\AA. The expanded $c$-axis, compared to pristine 3R-NbSe\textsubscript{2} \cite{brown1965layer, kalikhman1973transition}, indicates increased interlayer spacing, potentially weakening interlayer interactions and influencing the sample’s electronic and magnetic properties. EDAX analysis across various surface points yields average atomic percentages of 24\% Sn, 27\% Nb, and 49\% Se. FIG.~1.~(b) shows an HRTEM image of SnNbSe\textsubscript{2}, revealing a highly ordered lattice with atomic resolution, confirming excellent crystallinity. The atomic arrangement aligns with the $R\overline{3}m$ (166) space group symmetry. A simulated structural overlay matches the observed lattice along the $ab$-plane, as seen in FIG.~1.~(b). Further, SAED analysis (FIG.~1.~(c)) displays sharp diffraction spots, confirming high crystallinity. The indexed planes, (110), (010), (011), (014), and (114) support the rhombohedral structure.\par
  
  Magnetic properties of the SnNbSe\textsubscript{2} sample were characterized by susceptibility measurements as a function of temperature and the applied magnetic field. The zero field cooled warming (ZFCW) and field cooled warming (FCW) curves (FIG.~2.~(a)) show a sharp drop in susceptibility at $T_c = 4$~K, indicating a superconducting transition. The inset of FIG.~2 .~(a) displays an $M–H$ loop at 2~K, showing typical hysteresis behavior of type-II superconductors \cite{flukiger2012overview}. A small dip at 0.01~T, seen in the reducing field regime, is due to a flux jump due to sudden vortex rearrangement \cite{chikurov2021magnetic}. FIG.~2.~(b) presents $M$ vs. $H$ curves for 6R-SnNbSe\textsubscript{2} at various temperatures below $T_c$. At low fields, magnetization decreases linearly, reflecting strong diamagnetism and full flux expulsion. Deviations from linearity mark the onset of vortex penetration, defining the lower critical field $H_{c1}$ \cite{abrikosov2004nobel}. The inset of FIG.~2.~(b) plots \( T/T_c \) vs. reduced temperature \( T/T_c \), fitted with the model \( H_{c1}(T) = H_{c1}(0) [1 - (T/T_c)^2] \), yielding \( H_{c1}(0) \) = 1.76~mT from extrapolation of the data shown in FIG.~2.~(b). \par

 \begin{figure*}
\includegraphics[width=1\textwidth,height=6cm]{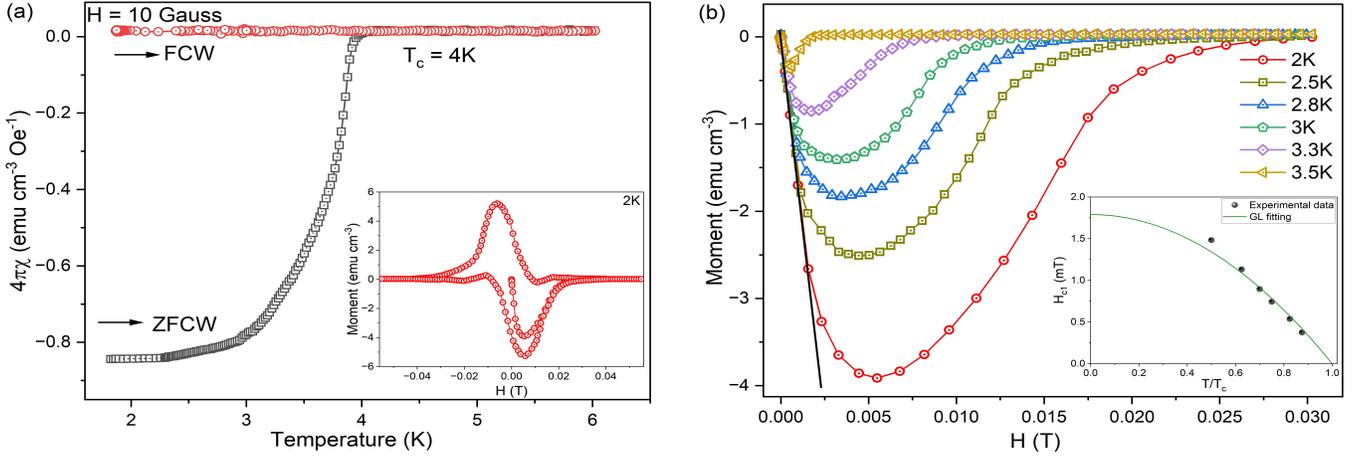} 
  \caption{(a) DC magnetization (ZFC and FC) with applied
10 Gauss magnetic field, and the inset shows $M-H$ loop at 2~K. (b) First quadrant of $M-H$ loop at different temperatures with linear interpolation for \( H_{c1} \) and inset shows extrapolated GL fitting of \( H_{c1} \).}
\end{figure*}

  Superconductivity is observed in 6R-SnNbSe\textsubscript{2} (space group 166, $R\overline{3}m$), but absent in pristine 3R-NbSe\textsubscript{2} (space group 160, $R3m$). This emergence is due to Sn intercalation, which induces structural and electronic changes. XRD and TEM reveal an expansion of van der Waals gaps, increasing the $c$-axis from $18.88$~\AA~in 3R-NbSe\textsubscript{2} to $27.83$~\AA~in 6R-SnNbSe\textsubscript{2}, thereby weakening interlayer coupling and enhancing two-dimensional (2D) superconductivity \cite{somoano1971superconductivity, zhang2022enhanced}. This decoupling also suppresses charge density wave (CDW) order \cite{lin2020patterns, yang2018enhanced}, which competes with superconductivity, thus favoring the superconducting phase. Strong intralayer interactions enhance electron-phonon coupling, which promotes Cooper pair formation \cite{mano2020straintronic}. Together, these effects stabilize superconductivity. Similar behavior is seen in Cu\textsubscript{x}TiSe\textsubscript{2} \cite{morosan2006superconductivity} and Sn-doped TaS\textsubscript{2} \cite{feig2020anisotropic}, where structural and chemical alteration suppress competing phases, enhancing superconductivity. \par

 FIG.~3.~(a) shows the temperature-dependent resistivity of SnNbSe\textsubscript{2}, measured using the linear four-probe method using a 10~mA DC current in the absence of an external magnetic field. Inset (i) confirms a superconducting transition at 4~K. The residual resistivity ratio $(RRR)$, calculated as $\textit{RRR} = \rho(300,\text{K})/\rho(0,\text{K})$, is 12. Above $T_c,$ the sample shows metallic behavior. Inset (i) of FIG.~4.~(a) reveals a linear resistivity trend above 100~K with a slope of $1.4 \times 10^{-2}~\, \text{m}\Omega\,\text{cm K}^{-1}$. This linear behaviour is a characteristic of electron-phonon scattering in metals \cite{naito1982electrical}. At lower temperatures, resistivity follows \(\rho(T) = \rho_0 + A T^2\), indicating the dominance of electron-electron scattering \cite{kusmartseva2009pressure}. Magnetoresistance measurements were performed under varying magnetic fields at 10 mA. As shown in FIG.~3.~(b), the superconducting transition temperature decreases with increasing field. The upper critical field ($H_{c2}$) values were extracted from resistivity–temperature curves. \par

 To further analyze the critical field behavior, the \(H_{c2}\) data were fitted to the equation:

\begin{equation}
H_{c2}(T) = H_{c2}(0)[1 - (T/T_c)^2]/[1 + (T/T_c)^2]
\end{equation}

This equation is aligned with the Ginzburg-Landau theory for a conventional single-band superconductor \cite{woollam1974positive}. However, the resulting fit, represented by the green dashed line in the inset (i) of FIG.~3.~(b), reveals a significant deviation from the experimental data.\par 

To achieve a more accurate representation, the two-band model was used to fit the experimental data. This model accounts for contributions from two superconducting bands with different electron-phonon coupling strengths. Application of the two-band model to \(H_{c2}(T)\) for SnNbSe$_2$ demonstrates excellent agreement with the experimental observations, as shown in the inset (i) of FIG.~3.~(b). The two-band model is formulated as follows \cite{gurevich2003enhancement}:
\begin{align}
    a_0 [\ln t + U(h)][\ln t + U(\eta h)] &+ a_1 [\ln t + U(h)] \notag \\
    &+ a_2 [\ln t + U(\eta h)] = 0
\end{align}

where, $h=\frac{H D_1}{2 \varphi_0 T}$, $\eta=\frac{D_2}{D_1}$,
$t= \frac{T}{T_c}$ and $(U(x) = \psi(x + 1/2) - \psi(1/2))$.
The coefficients are defined as; $a_0 =\frac{2w}{\lambda_0}$,  $a_1 = 1 + \frac{\lambda_-}{\lambda_0}$ and $a_2 = 1 - \frac{\lambda_-}{\lambda_0}$ with $w = \lambda_{11}\lambda_{22} - \lambda_{12}\lambda_{21}$ and $\lambda_0 = \sqrt{\lambda_-^2 + 4\lambda_{12}\lambda_{21}}$.

In these equations, \(\psi(x)\) denotes the digamma function, and \(\varphi_0\) is the magnetic flux quantum. The parameters \(\lambda_{ii}\) and \(\lambda_{ij}\) (for \(i \neq j\)) correspond to the intraband and interband electron coupling constants, while $D_1$ and $D_2$ are the electron diffusivities in bands 1 and 2, respectively.\par

The fitting procedure yielded the following parameters: \(H_{c2}(0) = 104.4 \, ~\text{mT}\), \(\lambda_{11} = 0.04\), \(\lambda_{22} = 4.3\), and \(\lambda_{12} = \lambda_{21} = 0.18\), with \(\eta = 10\).\par

The obtained superconducting coupling parameters provide essential information about the nature of superconductivity in SnNbSe$_2$. These parameters not only determine the existence of two-band superconductivity but also indicate the relative contribution of each band to the superconducting state. The values of $\lambda_{ii}$, referred to as intraband couplings, represent the magnitude of electron-phonon interaction in each band. The strong contrast between $\lambda_{11}$ and $\lambda_{22}$ values indicates that band 2 possesses a much stronger superconducting pairing interaction than band 1. The small interband coupling constants $\lambda_{ij}$ indicate minimal scattering between the two bands. Due to this weak coupling, the bands remain largely independent, resulting in two distinct superconducting energy gaps. The presence of such multiple gaps is a characteristic feature of multiband superconductors, similar to those observed in MgB$_2$ and other TMDs \cite{iavarone2002two, luo2020possible}.\par

The parameter $\eta = D_2/D_1$ describes the ratio of diffusivities between the two bands. A value of $\eta = 10$ indicates that electrons in band 2 experience significantly higher diffusivity compared to band 1.\par

\begin{figure*}
\includegraphics[width=1\textwidth,height=6cm]{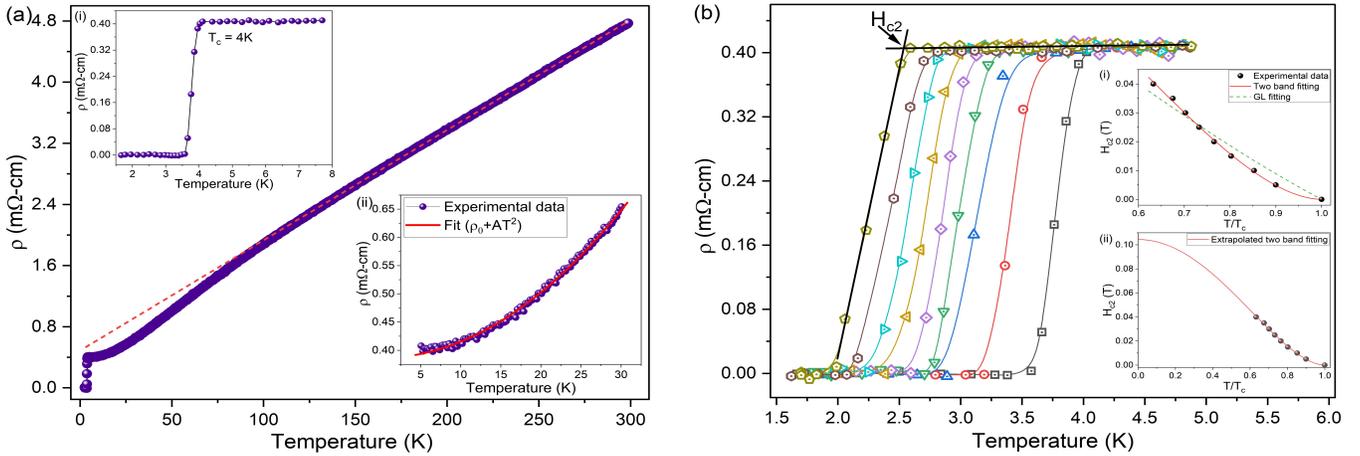} 
  \caption{(a) Zero field resistivity data over the range of temperature. Inset (i) shows the superconducting transition at 4~K, and inset (ii) shows the quadratic fitting at a lower temperature. (b) Temperature-dependent resistivity measured under magnetic fields ranging from 0 to 400 Gauss, in steps of 50 Gauss. Inset (i) shows Ginzburg-Landau and two-band model fittings of the upper critical field \( H_{c2}(T) \), and inset (ii) shows an extrapolated two-band fitting of \( H_{c2} \).}
\end{figure*}
   
  The upper critical field $H_{c2}(0)$ value of 104.4 mT is derived from the two-band fit model as shown in the inset (ii) of FIG.~3.~(b). This allows us to estimate the Ginzburg-Landau (GL) coherence length $\xi_{GL}$ using the standard GL relation \cite{tinkham2004introduction}:
\begin{equation}
H_{c2}(0) = \frac{\phi_0}{2 \pi \xi_{GL}^2}
\end{equation}

where \( \phi_0 = \frac{h}{2e} \approx 2.07 \times 10^{-15} \, \text{Wb} \) is the magnetic flux quantum. Solving for \( \xi_{GL} \) yields:

\[
\xi_{GL} = \sqrt{\frac{\phi_0}{2 \pi H_{c2}(0)}} \approx \sqrt{\frac{2.07 \times 10^{-15}}{2 \pi \times 0.1044}} \approx 562 \,~\text{Å}
\]

To further characterize the superconducting properties, we estimate the penetration depth \( \lambda_{GL} \) based on the lower critical field \( H_{c1}(0) \approx 1.76~\text{mT} \) \cite{tinkham1996books}.

\begin{equation}
H_{c1}(0) = \frac{\phi_0}{4 \pi \lambda_{GL}^2} \ln \left( \frac{\lambda_{GL}}{\xi_{GL}} \right)
\end{equation}

This yields:

\[
\lambda_{GL} \approx \sqrt{\frac{\phi_0 \ln (\lambda_{GL} / \xi_{GL})}{4 \pi H_{c1}(0)}}
\]

By substituting \( \xi_{GL} = 562 \,~\text{Å} \) and solving iteratively, we find \( \lambda_{GL} \approx 4385 \,~\text{Å} \).

With \( \lambda_{GL} \) and \( \xi_{GL} \) determined, we calculate the Ginzburg-Landau parameter \( \kappa_{GL} = \lambda_{GL} / \xi_{GL} \):

\[
\kappa_{GL} = \frac{4385}{562} \approx 7.80
\]

This value supports the classification of SnNbSe$_2$ as a type-II superconductor with \( \kappa_{GL} > 1/\sqrt{2} \approx 0.707 \).

For a better understanding of the superconducting pairing symmetry in 6R-SnNbSe$_2$, a comprehensive investigation of the temperature-dependent magnetic penetration depth, \( \Delta\lambda(T) \), was done using a tunnel diode oscillator (TDO) technique. This method offers a good sensitivity to changes in the resonant frequency in the superconducting state. The measured shift in resonant frequency, \( \Delta F \), was converted into a corresponding change in penetration depth, \( \Delta\lambda(T) \) using a calibration factor, \( G = 2.27 \,~\text{\AA/Hz} \) for our system \cite{yadav2024order}. The resulting penetration depth data, presented in FIG.~4.~(a), show a sharp and well-defined rise near the superconducting transition temperature \( T_c \approx 4.1 \,~\text{K} \), indicative of a high-quality sample and a clean superconducting transition. The low-temperature behavior of \( \Delta\lambda(T) \) was analyzed in detail to explore the superconducting pairing mechanism. In this regime, thermal fluctuations and vortex dynamics are minimal, making it ideal for probing the intrinsic pairing symmetry. \par

First, the experimental data were fitted using a conventional s-wave model derived from Bardeen-Cooper-Schrieffer (BCS) theory. A s-wave coupling is referred to as the formation of Cooper pairs with zero angular momentum $(l = 0)$ \cite{nikolic2011cooper}. This pairing symmetry results in an isotropic superconducting gap, which means the gap function $\Delta(\mathbf{k})$ remains constant across all directions on the Fermi surface \cite{scalapino2002bcs}. Thus it can be written as $\Delta(\mathbf{k})=\Delta_0$. Because of this directional uniformity, the s-wave gap is symmetric, which is the fundamental characteristic of conventional superconductors described by BCS theory. In momentum space, this isotropic nature leads to a spherical distribution of the superconducting gap in the Fermi surface \cite{schofield2009there}.\par

According to the BCS model, the low-temperature variation in penetration depth follows an exponential temperature dependence given by \cite{muhlschlegel1959thermodynamischen}:
\[
\Delta\lambda(T) = \lambda(0) \sqrt{\frac{\pi \Delta_0}{2k_B T}} \exp\left(-\frac{\Delta_0}{k_B T}\right),
\]
where \( \Delta_0 \) is the superconducting energy gap at zero temperature and \( \lambda(0) \) is the zero-temperature penetration depth. Here, the used value of  $\lambda(0)$ was obtained from upper critical fields. The data were fitted using the standard BCS model that yields a gap ratio \( \Delta_0 /k_BT_c = 1.47 \pm 0.12\). This value is significantly smaller than the values typically found for a weak-coupling s-wave superconductor in the BCS theory (\( \Delta_0 /k_B T_c = 1.76\)) \cite{mitra2017probing} as shown in FIG.~4.~(b), which doesn't give a good fit. The better fit yields a smaller gap ratio, suggests deviations from conventional BCS behavior, and may indicate the presence of anisotropic superconductivity or nodes in the gap function. These findings suggest that the material may exhibit an unconventional superconductivity \cite{tee2012penetration}.  Moreover, as shown in FIG.~4.~(b), this exponential model did not fit the experimental data well at low temperatures, as the curve of the model doesn’t match the actual data very well, especially at lower temperatures. These findings imply that the assumption of a uniform, fully gapped superconducting state is not valid for SnNbSe$_2$. \par

 \begin{figure*}
\includegraphics[width=1\textwidth,height=9cm]{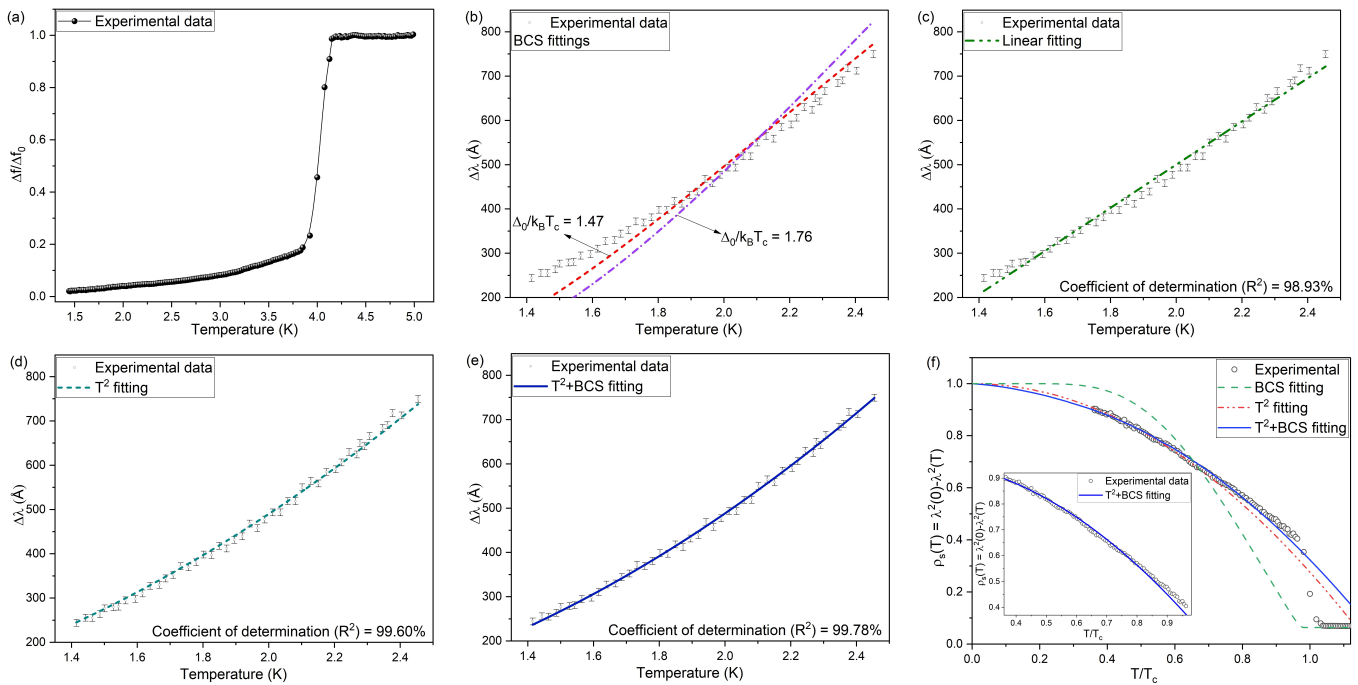}
\caption{(a) Temperature dependence of the normalized frequency shift \( \Delta f / \Delta f_0 \) in 6R-SnNbSe\(_2\) across the full measured temperature range. (b) Low-temperature variation of the penetration depth \( \Delta\lambda(T) \) in 6R-SnNbSe\(_2\), with two BCS fits: a standard fully gapped BCS model (dash-dotted line) and a BCS-like fit with a variable gap parameter (dashed line). (c) Linear fit of \( \Delta\lambda(T) \), (d) Quadratic (\( T^2 \)) fit, and (e) Combined fit using BCS theory and \( T^2 \) dependence in the low-temperature regime. (f) Superfluid density \( \rho_s(T) = \lambda^2(0)/\lambda^2(T) \) derived from \( \Delta\lambda(T) \), fitted over the entire temperature range, and the inset shows the fitted range focus on the combined fit of BCS and quadratic fit.}
\end{figure*}

To check if the superconducting gap might have nodes, the same data were also fitted with a linear temperature dependence, \( \Delta\lambda(T) \propto T \), which is usually expected in d-wave superconductors that have line nodes in the energy gap \cite{vandervelde2009evidence}. In d-wave superconductivity, the gap symmetry is characterized by a strong angular dependence, typically described by $\Delta(\mathbf{k}) = \Delta_0 (\cos k_x - \cos k_y)$, corresponding to $d_{x^2 - y^2}$ symmetry \cite{won2005bcs}. The pairing symmetry has an angular momentum $(l = 2)$ in the superconducting state. This type of gap changes between positive and negative values and drops to zero along certain directions in momentum space, which are called line nodes \cite{izawa2005line}. Specifically, the gap vanishes along the Brillouin zone diagonals $(k_x = \pm k_y)$, creating a four-lobed, butterfly-like shape in momentum space with alternating signs in adjacent lobes \cite{won2005bcs}. This anisotropic structure leads to low-energy quasiparticle excitations near the nodes and causes distinct physical behaviors such as a linear temperature dependence of the magnetic penetration depth at low temperatures \cite{bonalde2005low}. \par

The result of the linear fit is shown in the inset of FIG.~4.~(c), and it gives a slightly better match to the data, with \( R^2 = 98.93\% \). However, the fit still shows noticeable differences in the shape of the curve. This means SnNbSe$_2$ doesn't follow the linear behaviour that is typical for d-wave superconductors with well-defined nodes. \par

Next, the experimental data were fitted with a quadratic power-law behavior: $\Delta\lambda(T)~\propto~T^2$. The data fitting yielded a high coefficient of determination, $R^2 = 99.60\%$, indicating that the quadratic power-law model provides an excellent fit to our low-temperature penetration depth measurements (Fig.~4.~(d)). The $T^2$ dependence of the penetration depth at low temperatures is a signature of nodal point behavior in the superconducting gap structure \cite{teknowijoyo2018nodeless, smylie2022full}. This result suggests that SnNbSe$_2$ exhibits point nodes in its superconducting state. The presence of point nodes is often associated with unconventional pairing states such as p-wave superconductivity, in which the Cooper pairs form with odd parity and spin-triplet symmetry \cite{kaladzhyan2016characterizing, fu2010odd}. The odd parity means that the gap function satisfies \( \Delta(-\mathbf{k}) = -\Delta(\mathbf{k}) \) which leads to sign changes across the Fermi surface and often resulting in point nodes where the gap vanishes only at selected points in momentum space \cite{fu2014odd, venderbos2016odd}. These odd-parity pairing states are typically stabilized in systems with broken inversion symmetry or strong spin-orbit coupling \cite{kozii2015odd, wang2016topological}. These effects can give rise to topologically nontrivial superconducting phases that exhibit point nodes. We note that such quadratic fitting behavior is reported in other topological superconductors \cite{smylie2016evidence}. \par

Exploring the possibility that 6R-SnNbSe$_2$ exhibits mixed superconducting pairing symmetries, we applied a complex model that algebraically combines the BCS exponential behavior (indicative of a nodeless gap) with a quadratic power-law term (characteristic of nodal behavior). This combined model is motivated by theoretical descriptions of topological superconductors, in which the coexistence of nodeless and nodal pairing symmetries is expected \cite{zhu2023intrinsic, santos2010superconductivity, ghosh2010non}. Such scenarios often arise in systems that exhibit uncommon features like strong SOC and lack of inversion symmetry, both of which are present in 6R-SnNbSe$_2$.\par

The results of this fit are shown in Fig.~4.~(e). It achieves the highest coefficient of determination among all tested models, with $R^2 = 99.78\%$. This excellent agreement with experimental data suggests the superconducting state of 6R-SnNbSe$_2$ is best described by a juxtaposed gap structure that incorporates both gapped and gapless components. This observation is particularly important, as such a mixed pairing symmetry is a key signature of topological superconductivity and may provide the conditions for the emergence of Majorana fermions \cite{zhu2023intrinsic}. \par 

To further validate the possibility of the presence of a topological superconducting state, the normalized superfluid density \( \rho_s(T) = [\lambda(0)/\lambda(T)]^2 \) was determined using experimental data for the temperature-dependent magnetic penetration depth \cite{prozorov2000evidence}. The variation of \( \rho_s \) with reduced temperature \( T/T_c \) is plotted in Fig.~4.(f).

For comparison, the expected behavior of \( \rho_s(T) \) for a conventional isotropic s-wave superconductor was given by the relation \cite{tinkham2004introduction}:
\[
\rho_s(T) = 1 + 2 \int_{0}^{\infty} \frac{\partial f_0}{\partial E} \, d\varepsilon,
\]
where \( f_0 = [\exp(E/k_B T) + 1]^{-1} \) is the Fermi-Dirac distribution, and \( E = \sqrt{\varepsilon^2 + \Delta(T)^2} \) represents the quasi-particle energy. The temperature dependence of the superconducting gap \( \Delta(T) \) was reduced as \cite{gross1986anomalous}:
\[
\Delta(T) = \delta_{sc} k_B T_c \tanh \left[ \frac{\pi}{\delta_{sc}} \sqrt{a \left( \frac{\Delta C}{C} \right) \left( \frac{T_c}{T} - 1 \right)} \right],
\]
where \( \delta_{sc} = \Delta(0)/k_B T_c \), \( a = 2/3 \), and \(\Delta C/C \equiv C/\gamma T_c \). Keeping \( T_c = 4\)~K fixed, the obtained best fit has the following parameters: \( \Delta(0) = 0.808\,~\mathrm{meV} \), and \( C/\gamma T_c = 3.00 \pm 0.36 \). This is shown by the dashed green line in Fig.~4.(f). Clearly, the conventional BCS framework fails to match the experimental data. \par

A better description of the data is obtained by fitting with a quadratic power-law behavior, \(\rho_s(T) = 1-nT^2 \), as indicated by the red dashed-dotted line in Fig.~4.~(f). This model shows the low-temperature behavior more accurately and is consistent with the presence of nodal quasiparticles. This further supports the existence of point nodes in the superconducting gap, often associated with p-wave pairing symmetry.\par

In topological superconductors, where inversion symmetry is broken and spin-orbit coupling is strong, the superconducting state can involve a mixture of s-wave (nodeless) and p-wave (nodal) components. This mixed symmetry leads to a hybrid gap structure in which gapped and nodal regions coexist on the Fermi surface. The model, when fitted to the superfluid data, shows the best agreement across the full temperature range, as represented by the solid blue line in Fig.~4.~(f). The inset of Fig.~4.~(f), focusing on the intermediate temperature range (\(0.35 \leq T/T_c \leq 0.8\)), highlights the excellent consistency between the combined model and the experimental data.

These results provide strong evidence for the admixture of nodal and nodeless characteristics in the condensate pairing of 6R-SnNbSe\(_2\). Such features are associated with a topologically nontrivial superconducting state \cite{chia2003probing}.

\section{\label{sec:level1}CONCLUSION}
In summary, we have thoroughly studied the structural and electromagnetic properties of 6R-SnNbSe$_2$, a new polytype among transition metal dichalcogenides. The crystal structure of this new material is confirmed to be $R\overline{3}m$ through X-ray diffraction and transmission electron microscopy. Superconductivity was observed with a transition temperature of 4~K as verified by both resistive and magnetic measurements. The temperature dependence of the upper critical field follows a two-band model. Magnetic penetration depth measurements indicate that the superconducting order parameter symmetry is best described by an admixture of nodal and nodeless characteristics. Such odd parity triplet pairing is a sought-after signature of topological superconductivity. This conclusion is further supported by the temperature dependence of the superfluid density, \(\rho_s(T)\), which follows a model based on the combination of exponential and quadratic temperature dependencies. Overall, our results establish 6R-SnNbSe\(_2\) as a promising candidate for hosting topological superconductivity.

 \begin{acknowledgments}
    K Yadav acknowledges the Council for Scientific and Industrial Research (CSIR) for a Senior Research Fellowship. M Lamba thanks University
    Grant Commission (UGC) for financial support through SRF. We express our gratitude to the Department of Science and Technology, Government of India, for the low temperature and high field facility at JNU (FIST program) and the Nano-mission project (DST/NM/TUE/QM-10/2019(G)/6) for chemicals and consumables. We also thank the Advanced Instrumentation Research Facility (AIRF), JNU, for technical support.
\end{acknowledgments}

\bibliographystyle{plain} 
\bibliography{SnNbSe2}% Produces the bibliography via BibTeX.

\end{document}